\documentstyle [12pt] {article}
\input epsf
\newcommand{\gtwid}{\mathrel{\raise.3ex\hbox{$>$\kern-.75em\lower1ex
\hbox{$\sim$}}}}
\newcommand{\ltwid}{\mathrel{\raise.3ex\hbox{$<$\kern-.75em\lower1ex
\hbox{$\sim$}}}}
\newcommand{\beq}{\begin{equation}}
\newcommand{\eeq}{\end{equation}}
\newcommand{\beqs}{\begin{eqnarray}}
\newcommand{\eeqs}{\end{eqnarray}}

\catcode`@=11
\@addtoreset{equation}{section}
\@addtoreset{equation}{subsection}
\def\theequation{\ifnum\value{section}=0 \arabic{equation}\ignorespaces
\else \ifnum\value{section}=-1 A.\arabic{equation}\ignorespaces
\else \ifnum\value{subsection}=0 \thesection.\arabic{equation}\ignorespaces
\else \thesection.\arabic{subsection}.\arabic{equation}\ignorespaces
                           \fi
                      \fi
                 \fi}
\catcode`@=12

\begin{document}

\def\thefootnote{\fnsymbol{footnote}}
\baselineskip 6.0mm

\begin{flushright}
\begin{tabular}{l}
ITP-SB-96-49 
\end{tabular}
\end{flushright}

\vspace{4mm}
\begin{center}

{\bf Upper and Lower Bounds for Ground State Entropy of }

\vspace{2mm}

{\bf Antiferromagnetic Potts Models } 

\vspace{8mm}

\setcounter{footnote}{0}
Robert Shrock\footnote{email: shrock@insti.physics.sunysb.edu}
\setcounter{footnote}{6}
and Shan-Ho Tsai\footnote{email: tsai@insti.physics.sunysb.edu}

\vspace{6mm}

Institute for Theoretical Physics  \\
State University of New York       \\
Stony Brook, N. Y. 11794-3840  \\

\vspace{10mm}

{\bf Abstract}
\end{center}

    We derive rigorous upper and lower bounds for the ground state entropy of 
the $q$-state Potts antiferromagnet on the honeycomb and triangular lattices. 
These bounds are quite restrictive, especially for large $q$. 

\vspace{16mm}

\pagestyle{empty}
\newpage

\pagestyle{plain}
\pagenumbering{arabic}
\renewcommand{\thefootnote}{\arabic{footnote}}
\setcounter{footnote}{0}

    Nonzero ground state disorder and associated entropy, $S_0 \ne 0$, is an 
important subject in statistical mechanics; a physical realization is 
provided by ice, for which 
$S_0 = 0.82 \pm 0.05$ cal/(K-mole), i.e., $S_0/k_B = 0.41 \pm 0.03$ 
\cite{ice,liebwu}.  Ground state (g.s.) entropy may or may not be associated
with frustration. An early example with frustration is the Ising (equivalently,
$q=2$ Potts) antiferromagnet on the triangular lattice \cite{wannier}.
However, g.s. entropy is also exhibited in the simpler context of models
without frustration, such as the $q$-state Potts antiferromagnet (AF) 
\cite{potts}-\cite{baxterbook} on the square ($sq$) and honeycomb ($hc$)
lattices for (integral) $q \ge 3$ and on the triangular ($tri$) lattice for 
$q \ge 4$.  Of these three 2D lattices, $S_0$ has been calculated exactly 
for the triangular case \cite{baxter87}, but, aside from the single value 
$S_0(sq,q=3)/k_B=(3/2)\ln(4/3)$ \cite{lieb}, not for the square or honeycomb 
lattices.  Therefore, it is valuable to have rigorous upper and lower bounds on
this quantity.  Using a ``coloring matrix'' method, Biggs derived such 
bounds for the square lattice \cite{biggs77}.  Here we shall extend his 
method to derive analogous bounds for the honeycomb lattice and compare the 
results with our recent Monte Carlo measurements \cite{p3afhc,w1} and with 
large-$q$ series \cite{kewser}.  We also derive such bounds for the triangular
lattice; the interest in this case is that the bounds can be compared with the 
exact result of Baxter \cite{baxter87}. 

   We make use of the fact that the partition function at $T=0$, 
$Z(\Lambda,q,K=-\infty)$, for the $q$-state zero-field Potts AF on a lattice 
$\Lambda$ (where $K=\beta J$, $\beta=1/(k_BT)$, and $J<0$
denotes the spin-spin coupling) is equal to the chromatic polynomial
$P(\Lambda,q)$.  Here, $P(G,q)$ is the number of ways of coloring the graph $G$
with $q$ colors such that no adjacent vertices (sites) have the same color
\cite{rtrev}.  Define the reduced, per site free energy for the Potts AF in the
thermodynamic limit as 
$f(\Lambda,q,K) = \lim_{N \to \infty} N^{-1} \ln Z(\Lambda,q,K)$.  From the 
general relation between the entropy $S$, the internal energy $U$, and the 
reduced free energy, $S = \beta U + f$ (henceforth, we use units such
that $k_B = 1$), together with the property that 
$\lim_{K \to -\infty}\beta U(\beta) = 0$, as is true of the $q$-state Potts AF
models considered here, it follows that $S_0(\Lambda,q) = 
f(\Lambda,q,K=-\infty) = \ln W(\Lambda,q)$, where $W(\Lambda,q)$ is the 
asymptotic limit
\beq
W(\Lambda,q) = \lim_{N \to \infty} P(\Lambda,q)^{1/N}
\label{w}
\eeq
Given this connection, we shall express our bounds on the g.s. entropy
$S_0(\Lambda,q)$ in terms of the equivalent function $W(\Lambda,q)$.  As we 
have discussed earlier \cite{w1}, the formal eq. (\ref{w}) is not, in general,
adequate to define $W(\Lambda,q)$ because of a noncommutativity of
limits 
\beq
\lim_{N \to \infty} \lim_{q \to q_s} P(\Lambda,q)^{1/N} \ne
\lim_{q \to q_s} \lim_{N \to \infty} P(\Lambda,q)^{1/N}
\label{wnoncomm}
\eeq
at certain special points $q_s$.  We denote the definitions based on the 
first and second orders of limits in (\ref{wnoncomm}) as 
$W(\Lambda,q)_{D_{nq}}$ and $W(\Lambda,q)_{D_{qn}}$, respectively.  This 
noncommutativity can occur for $q < q_c(\Lambda)$, where $q_c(\Lambda)$ 
denotes the maximal (finite) real value of $q$ where $W(\Lambda,q)$ is 
nonanalytic \cite{w1}.  Since 
$q_c(hc)=(3+\sqrt{5})/2 = 2.618...$ and $q_c(tri)=4$ (and 
$q_c(sq)=3$) \cite{w1}, it follows that for the ranges of
$q$ considered here, viz., (real) $q \ge 3$ for $\Lambda=hc,sq$ and $q \ge 4$
for $\Lambda=tri$, one does not encounter the noncommutativity
(\ref{wnoncomm}). 

  To proceed, we consider a sequence of (regular) $m \times n$ \ \ 
2D lattices of
type $\Lambda$, with periodic boundary conditions (PBC's) in both directions, 
and $m$ and $n$ even to maintain the bipartite property for finite square and 
honeycomb lattices and thereby avoid frustration.  
For $\Lambda=sq$, Biggs introduced the notion of a coloring matrix $T$, 
somewhat analogous to the transfer matrix for statistical mechanical spin 
models.  The construction of $T$ begins by considering an $n$-vertex circuit 
$C_n$ along a column of $\Lambda$, i.e., a ring around the toroidal
lattice, given the PBC's.  The number of allowed $q$-colorings of this circuit
is $P(C_n,q) = (q-1)[(q-1)^{n-1} + (-1)^n]$.  
Now focus on two adjacent columnar circuits, 
$C_n$ and $C_n'$.  Define compatible $q$-colorings of these adjacent circuits
as colorings such that no two horizontally adjacent vertices on the circuits
have the same color.  One can then associate with this pair of adjacent
columnar circuits an ${\cal N} \times {\cal N}$ dimensional symmetric matrix 
$T$, where ${\cal N} = P(C_n,q)=P(C_n',q)$ 
with entries $T_{C_n,C_n'} = T_{C_n',C_n}=$ 1 or 0 if the 
$q$-colorings of $C_n$ and $C_n'$ are or are not compatible, respectively. 
Then for fixed $m,n$, $P(\Lambda_{m \times n},q) = Tr(T^m)$.  For a given $n$,
since $T$ is a nonnegative matrix, one can apply the Perron-Frobenius 
theorem \cite{pf} to conclude that $T$ has a real positive maximal eigenvalue 
$\lambda_{max,n}(q)$.  Hence, for fixed $n$, 
\beq
\lim_{m \to \infty} Tr(T^m)^{1/(mn)} \to \lambda_{max}^{1/n}
\label{trlim}
\eeq
so that 
\beq
W(\Lambda,q) = \lim_{n \to \infty} \lambda_{max}^{1/n}
\label{wlim}
\eeq
Denote the column sum $\kappa_j(T) = \sum_{i=1}^{\cal N} T_{ij}$ (equal to 
the row sum $\rho_j(T)=\sum_{i=1}^{\cal N}T_{ji}$ since $T^T=T$) and 
$S(T) = \sum_{i,j=1}^{\cal N} T_{ij}$; note that $S(T)/{\cal N}$ is the average
row (column) sum.  Combining the bounds for a 
general nonnegative ${\cal N} \times {\cal N}$ matrix $A$, 
\cite{pf} 
\beq
\min\{\gamma_j(A) \} \le \lambda_{max}(A) \le \max\{\gamma_j(A) \} \ , 
\quad {\rm for} \quad \gamma_j = \kappa_j \quad {\rm or} \quad \rho_j
\label{genuplowbound}
\eeq
with the ($k=1$ case of the ) more restrictive lower bound applicable to a 
symmetric nonnegative matrix \cite{london}, 
\beq
\biggl [\frac{S(T^k)}{\cal N} \biggr ]^{1/k} \le \lambda_{max} \ , \quad 
{\rm for} \quad k=1, 2, ... 
\label{lbound}
\eeq
we have
\beq
\frac{S(T)}{\cal N} \le \lambda_{max}(T) \le \max\{\kappa_j(T)\}
\label{uplowbound}
\eeq

   For the honeycomb lattice, we find that the analogue of the circuit $C_n$ on
the square lattice is the set of vertical dimers shown in Fig. 1(a), which we
denote as $p$.  With $m$ and $n$ even to maintain a bipartite lattice,
there are $n/2$ dimers in $p$, and the total number of $q$-colorings of
these dimers is $N_{hc,n}=[q(q-1)]^{n/2}$.  We next associate the $T$ matrix
$T(hc_n)$ with two adjacent sets of dimers $p$ and $p'$ (see Fig. 1(a));
$T(hc_n)$ is thus a $N_{hc,n} \times N_{hc,n}$ matrix.  
Two $q$-colorings of the
dimer sets $p$ and $p'$ are compatible if and only if the horizontally
adjacent vertices have different colors, and $T_{p,p'} =1$ or 0, respectively,
if these colorings are compatible or incompatible.   We observe that $S(T) =
P(C_{2n},q)$. Therefore, 
\beq
\frac{S(T(hc_n))}{N_{hc,n}} = 
\frac{(q-1)\Bigl [ (q-1)^{2n-1} + 1 \Bigr ] }{ [q(q-1)]^{n/2}}
\label{hclow}
\eeq

\begin{figure}
\vspace{-3 cm} 
\epsfxsize=3.5in
\begin{center}
\leavevmode
\epsffile{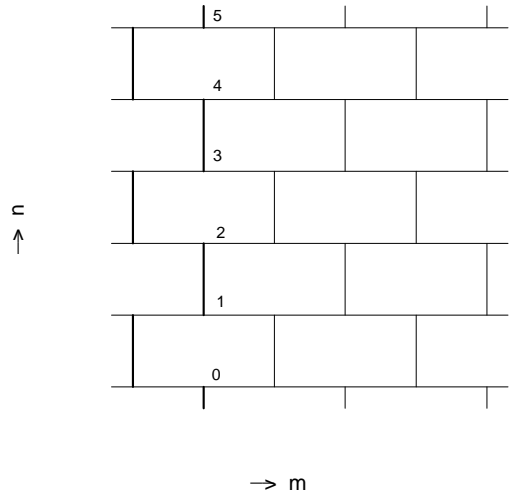}
\end{center}
\vspace{-5 cm}
\epsfxsize=3.5in
\begin{center}
\leavevmode
\vspace{-2 cm}
\epsffile{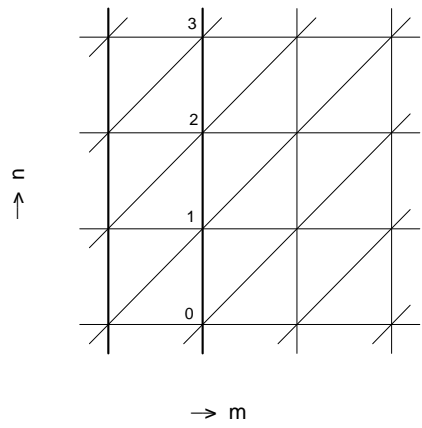}
\end{center}
\caption{(a) Honeycomb and (b) triangular lattices. See text for discussion.}
\label{latticefig}
\end{figure}

To calculate
the maximal column sum, we consider two neighboring sets of dimers 
$p$ and $p'$, with $n$ sites each labeled by $i=0,1,...,n-1$ (see Fig.1(a)). 
Let the sites of set $p$ be colored in such a way that sites on the same dimer
have different colors (choosing one such configuration of colors corresponds 
to fixing one column in the color matrix $T$). Let $X_j$ denote the 
number of $q$-colorings of sites $i=0$ to $i=j$ of $p'$, such that a site $i$
in set $p$ has a different color from a site $i$ on $p'$. If $j$ is odd, the
coloring of the $j$'th site in $p'$ is only constrained to be different from 
the coloring of the adjacent $j$'th site in $p$, so
$X_j=(q-1)X_{j-1}$. \cite{f1}  The
color assigned to an even-$j$ site in $p'$ must be different from the color 
of (i) the other member of the dimer in $p'$ and (ii) the adjacent $j$'th 
site in $p$; hence, $X_j=(q-2)X_{j-1}+Y_{j-1}$, where $Y_{j-1}$ denotes the 
number of colorings for which site $j$ of $p$ has the same color as site 
$j-1$ of $p'$. Note that $Y_{j-1}$ is a subset of $X_{j-2}$, i.e. 
$Y_{j-1}\le X_{j-2}$. Thus
\beq
X_j \le (q-2)X_{j-1} + X_{j-2}, \qquad {\rm for \ \  even} \ \  j, \qquad
 2 \le j \le n-2 
\label{xjeven}
\eeq
\beq
X_j=(q-1)X_{j-1} \ , \qquad {\rm for \ \  odd} \ \ j, \qquad 1 \le j \le n-1
\label{xjodd}
\eeq
Using eq.(\ref{xjodd}) in eq.(\ref{xjeven}) and setting $j=2\ell$, we have 
\beq
X_{2\ell} \le \Bigl [ (q-1)(q-2)+1 \Bigr ]X_{2\ell-2} \ ,\qquad {\rm for} \quad
2 \le 2\ell \le n-2 \ ,
\eeq
which yields $X_{2\ell} \le \Bigl [ (q-1)(q-2)+1 \Bigr ]^\ell X_0$, where 
$X_0=q-1$.
It follows that 
\beq
X_{n-1}=(q-1)X_{n-2} \le (q-1)^2 \Bigl [ (q-1)(q-2)+1 \Bigr ]^{(n-2)/2}
\label{xnmin1}
\eeq
Because $\max \{ \kappa_j(T(hc_n) \} \le X_{n-1}$, we obtain 
\beq
\max \{\kappa(T(hc_n)) \} \le (q-1)^2 [q^2-3q+3]^{(n-2)/2}
\label{hcup}
\eeq
Hence, using (\ref{trlim}) and (\ref{wlim}), we derive the bounds 
\beq
\frac{(q-1)^{3/2}}{q^{1/2}} \le W(hc,q) \le (q^2-3q+3)^{1/2}
\quad {\rm for} \quad q \ge 3
\label{whcb} 
\eeq
The bounds are also seen to apply for the case $q=2$ if one uses as a
definition $W(hc,2) \equiv W(hc,2)_{D_{nq}}=1$ given by the first order of
limits in eq. (\ref{wnoncomm}). 

Similarly, for the triangular lattice we define the color matrix $T(tri_n)$
by considering the compatibility of $q$-colorings of two neighboring $n$-vertex
circuits $C_n$ and $C_n'$. An example of such adjacent circuits is shown
by the darker lines in Fig. 1(b). Here $T(tri_n)$ is a $N_{tri,n} \times
N_{tri,n}$ matrix, where $N_{tri,n}=P(C_n,q)=P(C_n',q)$.  
For the triangular lattice with periodic or open boundary conditions in the
vertical direction, $S(T)$ is equal, respectively, to the numbers 
$P(cctri_n,q)$ and $P(octri_n,q)$ of $q$-colorings of a cyclic or open chain 
of triangles with $2n$ vertices.  In the $n \to \infty$ limit of interest here,
$\lim_{n \to \infty}P(cctri_n,q)^{1/n} = 
 \lim_{n \to \infty}P(octri_n,q)^{1/n}$, so it does not matter which type of
chain we use.  An elementary calculation yields
\beq
P(octri_n,q)=q(q-1)(q-2)^{2n-2}
\label{poctrin}
\eeq
so
\beq
\lim_{n \to \infty} \biggl [\frac{S(T(tri_n))}{N_{tri,n}}\biggr ]^{1/n} = 
\frac{(q-2)^2}{q-1}
\label{stri}
\eeq

    To calculate $\max \{ \kappa_j(T(tri_n)) \}$, we derive, as before, 
an upper bound for $X_{n-1}$. Each vertex of
$C'_n$ is connected to two vertices of $C_n$, hence each of the $X_{j-1}$
colorings of the sites $i=0$ to $i=j-1$ of the circuit $C'_n$ can be
extended in at least $q-3$ ways to the site $i=j$. Thus, for the triangular
lattice, the equivalent of eqs. (\ref{xjeven}) and (\ref{xjodd}) is
\beq
X_j \le (q-3)X_{j-1}+X_{j-2}, \qquad 2 \le j \le n-1,
\label{xjtri}
\eeq
with $X_0=q-2$ and $X_1 \le (q-3)(q-2)+1$ \ \cite{f2}. 
This recursion relation is of
the same form as the one obtained previously \cite{biggs77} for the square
lattice, and can thus be solved by the same method.    
We thus find for the triangular lattice the inequality
\beq
\frac{(q-2)^2}{q-1} \le W(tri,q) \le \frac{1}{2} \Bigl [ q-3 +
(q^2-6q+13)^{1/2} \Bigr  ] \qquad {\rm for} \qquad q \ge q_c(tri)=4
\label{wtrib}
\eeq
(For $q=3$, the ground state entropy is zero, i.e., $W(tri,3)_{D_{nq}}=1$.) 

   Denote the lower and upper bounds for these lattices as 
$W(\Lambda,q)_{\ell}$ and $W(\Lambda,q)_{u}$, respectively.  We observe that
as $q$ increases, these bounds rapidly approach each other, and hence restrict 
the exact values $W(\Lambda,q)$ very accurately.  This can be seen as a
consequence of the fact that, aside from the obvious prefactor $q$, 
$W(\Lambda,q)_\ell$ and $W(\Lambda,q)_u$ are the same up to $O(q^{-2})$:
\beq
q^{-1}W(hc,q)_{\ell} = 1 - \frac{3}{2}q^{-1} + \frac{3}{8}q^{-2} + 
\frac{1}{16}q^{-3} + O(q^{-4})
\label{wchlowtay}
\eeq
\beq
q^{-1}W(hc,q)_u = 1 - \frac{3}{2}q^{-1} + \frac{3}{8}q^{-2} +
\frac{9}{16}q^{-3} + O(q^{-4})
\label{wchuptay}
\eeq
and similarly with $q^{-1}W(tri,q)_b$, $b=\ell,u$. (This was also true for the
$\Lambda=sq$ bounds \cite{biggs77}).

   In Table 1, we compare the bounds (\ref{whcb}) for $\Lambda=hc$ with our 
recent Monte Carlo measurements of $W(hc,q)$ \cite{p3afhc,w1}.  We also compare
our bounds (\ref{wtrib}) for $\Lambda=tri$ with the exactly known results of
Baxter \cite{baxter87}.  For reference, Table 1 includes a similar comparison 
of the $\Lambda=sq$ bound with the known $q=3$ value
\cite{lieb} and Monte Carlo measurements \cite{chenpan,bakaev,w1} for $q > 3$. 
We see that as $q$ increases past $q \simeq 4$, the upper and lower bounds 
bracket the actual respective values quite closely, and that the latter 
values lie closer to the lower bounds. 

\begin{table}
\begin{center}
\begin{tabular}{|c|c|c|c|c|c|c|} \hline \hline & & & & & & \\
$q$ & $\frac{W(hc,q)_\ell}{W(hc,q)_{MC}}$ & $\frac{W(hc,q)_u}{W(hc,q)_{MC}}$ & 
      $\frac{W(sq,q)_\ell}{W(sq,q)}$ & $\frac{W(sq,q)_u}{W(sq,q)}$ &
      $\frac{W(tri,q)_\ell}{W(tri,q)}$    & $\frac{W(tri,q)_u}{W(tri,q)}$ \\
& & & & & & \\
\hline \hline
3  &  0.98390(60)    &  1.04358(65)   &  0.97425(55)  &  1.05091(60)  
   &  $-$    &  $-$ \\
4  &  0.99781(60)    &  1.01612(60)   &  0.99844(65)  &  1.03305(65)  
   &  0.91262  &  1.107485 \\
5  &  0.99948(55)    &  1.00726(55)   &  0.99970(60)  &  1.01593(60)
   &  0.99377        &  1.06630  \\
6  &  0.99978(65)    &  1.00377(65)   &  0.99992(60)  &  1.00851(60)
   &  0.99879        &  1.03087  \\
7  &  0.99988(65)    &  1.00220(65)   &  0.99996(60)  &  1.00498(60)
   &  0.99963        &  1.01628  \\
8  &  0.99999(60)    &  1.00145(60)   &  0.99996(65)  &  1.00312(65)
   &  0.99986        &  1.00953  \\
9  &  1.00001(60)    &  1.00099(60)   &  0.99995(65)  &  1.00206(65)
   &  0.99994        &  1.00602  \\
10 &  0.99994(60)    &  1.00063(60)   &  0.99986(60)  &  1.00134(60)
   &  0.99997        &  1.00404  \\
\hline
\end{tabular}
\end{center}
\caption{Comparisons of lower and upper bounds with Monte Carlo measurements 
of $W(hc,q)$ and exact values of $W(tri,q)$, respectively, for 
$q_c(\Lambda) \le q \le 10$.  An analogous comparison is included for
$\Lambda=sq$.} 
\label{wtable}
\end{table}

   To understand why the actual values of $W(\Lambda,q)$ lie closer to the 
respective lower bounds, 
we compare the large-$q$ series with the expansions of these lower bounds.  
For a lattice $\Lambda$ with coordination number $\zeta$, the large $q$ series
can be written in the form 
\beq
W(\Lambda,q) = q(1-q^{-1})^{\zeta/2} \; \overline W(\Lambda,y)
\label{wseriesdef}
\eeq
where $\overline W(\Lambda,y)=1+\sum_{n=1}^\infty w_n y^n$ with 
$y = \frac{1}{q-1}$. 
Defining the analogous functions $\overline W(\Lambda,y)_b$ via 
\beq
W(\Lambda,q)_b = q(1-q^{-1})^{\zeta/2} \; \overline W(\Lambda,y)_b \ , \quad
b = \ell, \ u \ , 
\label{wbarb}
\eeq
we obtain $\overline W(hc,y)_\ell = 1$, 
which agrees to the first five terms, i.e., to order $O(y^4)$, with the 
series \cite{kewser} $\overline W(hc,y) = 1 + y^5 + 2y^{11} + O(y^{12})$, while
$\overline W(hc,y)_u = 1 + y^3/2 + O(y^6)$. 
We also calculate $\overline W(tri,y)_\ell = (1-y^2)^2$, 
which agrees to the first five terms, i.e. to $O(y^4)$, with the series
expansion of the exact Baxter result, 
$\overline W(tri,y) = 1 -2y^2 + y^4 + y^5 + O(y^6)$, while $\overline
W(tri,q)_u = 1 -2y^2 + 2y^3 + O(y^4)$. Finally, 
$\overline W(sq,y)_\ell = 1 + y^3$, 
which agrees to the first seven terms, i.e, to $O(y^6)$, with the series
\cite{kewser,bakaev} $\overline W(sq,y) = 1 + y^3 + y^7 + O(y^8)$, while 
$\overline W(sq,y)_u = 1 + 2y^3 + O(y^4)$. 

   In summary, we have derived rigorous upper and lower bounds for the 
(exponent of the) ground state entropy of the Potts antiferromagnet on the 
honeycomb and triangular lattices and have shown that these are very
restrictive for large $q$.  Since nonzero ground state entropy sheds light on
some of the most fundamental properties of statistical mechanics, it is of
interest to derive similar bounds for other lattices; work on this is in 
progress. 

This research was supported in part by the NSF grant PHY-93-09888.

\vfill
\eject

\begin{thebibliography}{99}
\bibitem{f1}{The restriction on the coloring of site 0 in $p'$ will be taken
into account automatically by our use of periodic boundary conditions here.
Alternatively, one could use an open chain; these choices are equivalent in 
the $n \to \infty$ limit.}
\bibitem{f2}{For cases where the colors on the site $j=1$ and the site below 
$j=0$ on $C_n$ are different, the
expression for $X_1$ is realized as an equality; if these colors are the same,
then it is realized as an inequality.}
\end{thebibliography}
\end{document}